# EigenRec: Generalizing PureSVD for Effective and Efficient Top-N Recommendations

**Athanasios N. Nikolakopoulos · Vassilis Kalantzis · Efstratios Gallopoulos · John D. Garofalakis**

**Abstract** We introduce EigenRec; a versatile and efficient Latent-Factor framework for Top-N Recommendations that includes the well-known *PureSVD* algorithm as a special case. EigenRec builds a low dimensional model of an inter-item proximity matrix that combines a similarity component, with a scaling operator, designed to control the *influence of the prior item popularity* on the final model. Seeing PureSVD within our framework provides intuition about its inner workings, exposes its inherent limitations, and also, paves the path towards painlessly improving its recommendation performance. A comprehensive set of experiments on the `MovieLens` and the `Yahoo` datasets based on widely applied performance metrics, indicate that EigenRec outperforms several state-of-the-art algorithms, in terms of *Standard* and *Long-Tail* recommendation accuracy, exhibiting low susceptibility to sparsity, even in its most extreme manifestations – the *Cold-Start* problems. At the same time EigenRec has an attractive computational profile and it can apply readily in large-scale recommendation settings.

## 1 Introduction

Collaborative Filtering (CF) is commonly regarded as one of the most effective approaches to building Recommender Systems (RS). Given a set of users, a set of items and – implicitly or explicitly – stated opinions about how much a user likes or dislikes the items he has already seen, CF techniques try to build "neighborhoods", based on the similarities between users (*user-oriented* CF) or items (*item-oriented* CF) as depicted in the data, in order to predict

Athanasios N. Nikolakopoulos[1] and Vassilis Kalantzis[2]
[1]Digital Technology Center and [2]Department of Computer Science and Engineering, University of Minnesota, Minneapolis, MN, USA.
E-mail: {anikolak,kalan019}@umn.edu

Efstratios Gallopoulos and John D. Garofalakis
Department of Computer Engineering and Informatics, University of Patras, Greece. E-mail: {stratis,garofala}@ceid.upatras.gr

preference scores for the unknown user-item pairs, or provide a list of items that the user might find preferable.

Despite their success in real application settings, CF methods suffer from several problems that remain to be resolved. One of the most significant such problems arises from the insufficiency of available data and is typically referred to as the *Sparsity* problem [12]. Sparsity is known to impose severe limitations to the quality of recommendations [8], and to decrease substantially the diversity and the effectiveness of CF methods – especially in recommending unpopular items (*Long-Tail* problem) [42]. Unfortunately, sparsity is an innate characteristic of recommender systems since in the majority of realistic applications, users typically interact with only a small percentage of the available items, with the problem being intensified even more, by the fact that newcomers with no ratings at all are frequently added to the system (*Cold-Start* problem [8,31]).

While traditional neighbourhood-based CF techniques are very vulnerable to sparsity, *Graph-Based* methods manage to cope a lot better [12]. The fundamental characteristic that makes the methods of this family well-suited for alleviating problems related to *limited coverage* and sparsity is that they allow elements of the dataset that are not directly connected to "influence" each other by propagating information along the edges of an underlying graph [12]. Then, the transitive relations captured in this way can be used to recommend items either by estimating measures of proximity between the corresponding nodes [20,30] or by computing similarity scores between them [14].

However, despite their potential in dealing with sparsity, graph-based techniques usually exhibit poor scalability and heavy computational profile – a fact that limits their applicability in large-scale recommendation settings. *Latent Factor* methods, on the other hand, present a more viable alternative [12,18,24,32,37]. The fundamental premise behind using latent factor models for building recommender systems is that user's preferences are influenced by a set of "hidden taste factors" that are usually very specific to the domain of recommendation [32]. These factors are generally not obvious and might not necessarily be intuitively understandable. Latent Factor algorithms, however, can infer those factors by the user's feedback as depicted in the rating data. Generally speaking, the methods in this family work by projecting the elements of the recommender database into a denser subspace that captures their most meaningful features, giving them the power to relate previously unrelated elements, and thus making them less susceptible to sparsity [12].

**Motivation & Contributions.** A very simple and widely used latent factor algorithm for top-$N$ recommendations is *PureSVD* [11]. The algorithm considers all missing values in the ratings matrix as zeros, and produces recommendations by reconstructing it based on its truncated singular value decomposition[1]. Cremonesi *et al.* [11], after evaluating PureSVD's performance against various latent factor-based algorithms and neighbourhood models, found that it was able to achieve competitive top-$N$ recommendation performance compared to sophisticated matrix factorization methods [24,25] and other popular CF techniques. However, despite showing promising qualitative results and being fairly simple to apply, PureSVD as presented in [11] does not lend itself into fertile generalizations, nor does it leave room for qualitative improvements. The method is typically being used almost like an enigmatic "black box" that takes the ratings matrix as an input, and outputs its low-rank estimate – successfully perturbing, in the process, the previously zero values of the ratings matrix into something useful. But, *is there a more fruitful way to look at PureSVD? A way that can give more intuition about how it works and how it can be improved?*

[1] Note that even though the actual values of the reconstructed matrix do not have a meaning in terms of ratings, they induce an ordering of the items which is sufficient for recommending top-$N$ lists.

In this work[2], in an attempt to shed more light to these questions, we revisit the algorithmic inner-workings of PureSVD aiming to expose the "modeling scaffoldings" behind its mathematical structure. Interestingly, this approach provides an illustrative reformulation of the model that paves the path towards a straightforward generalization to a whole family of related methods – which we denote **EigenRec** – that can lead to qualitatively superior, and computationally attractive top-$N$ recommendation schemes.

- EigenRec works by building a low-dimensional subspace of a novel proximity matrix comprising *scaled inter-item similarity scores*. The pure similarity component can be defined by utilizing any reasonable measure one deems appropriate for the recommendation problem under consideration (here we use three standard similarity functions that were found to combine simplicity and effectiveness). The scaling component on the other hand, allows for fine-tuning the influence of the prior item popularity on the final proximity scores; a property that empowers our method to improve the produced recommendation lists significantly.
- One of our primary concerns pertains to the computability of our method in realistic big data scenarios. Our modeling approach implies immediate computational gains with respect to PureSVD, since it reduces the computation of a truncated singular value decomposition to the solution of a simpler symmetric eigenvalue problem applied to a linear operator of significantly lower dimensions. For problems that fit in a single machine, EigenRec can be computed readily using any off-the-shelf eigensolver. However, to ensure practical application of the method even for very large datasets, we propose a parallel approach for computing EigenRec based on a computationally efficient Krylov subspace procedure – namely the *Lanczos Method*. We discuss in detail its *parallel implementation* in distributed memory environments and we perform several tests using real-world datasets, thus ensuring the applicability of our method in large-scale scenarios[3].
- We conduct a comprehensive set of qualitative experiments on the `MovieLens` and `Yahoo` datasets and we show that even the simple members of the EigenRec family we are considering here, outperform several state-of-the-art methods, in widely used metrics, achieving high-quality results even in the considerably harder task of recommending *Long-Tail* items. EigenRec displays low sensitivity to the sparsity of the underlying space and shows promising potential in alleviating a number of related problems that occur commonly in recommender systems. This is true both in the very interesting case where sparsity is localized in a small part of the dataset – as in the *New Users* problem, and in the case where extreme levels of sparsity are found throughout the data – as in the *New Community* problem.

The rest of the paper is organized as follows: In Section 2.1, we revisit PureSVD and we "rediscover" it under different modeling lenses in order to set the intuitive grounds behind the EigenRec approach, which is then presented formally in Section 2.2. In Section 3, we present the EigenRec algorithm, we comment on its computational complexity and delve into the details behind its parallelization. The qualitative evaluation of EigenRec, including experimental methodology, metrics definition, a detailed discussion of the competing recommendation methods, as well as top-$N$ recommendation results in standard, long-tail and cold-start scenarios, are presented in Section 4. Our computational tests are presented in Section 5. Section 6 comments on related literature and, finally, Section 7 concludes this work.

[2] A preliminary version of this work has been presented in [29].

[3] High-level and MPI implementations of EigenRec can be found here: `https://github.com/nikolakopoulos/EigenRec`

## 2 EigenRec recommendation framework

**Notation.** All vectors are denoted by bold lower-case letters and they are assumed to be column vectors (e.g., $\mathbf{v}$). All matrices are represented by bold upper-case letters (e.g., $\mathbf{Q}$). The $j^{th}$ column and the $i^{th}$ row of matrix $\mathbf{Q}$ are denoted $\mathbf{q_j}$ and $\mathbf{q_i}^\intercal$, respectively. The $ij^{th}$ element of matrix $\mathbf{Q}$ is denoted as $Q_{ij}$. We use diag($\mathbf{Q}$) to refer to the matrix that has the same diagonal with matrix $\mathbf{Q}$ and zeros elsewhere, and diag($\mathbf{v}$) to denote the matrix having vector $\mathbf{v}$ on its diagonal, and zeros elsewhere. Furthermore, $\|\cdot\|$ denotes a norm that – unless stated otherwise – is assumed to be the Euclidean. We use calligraphic upper-case letters to denote sets (e.g., $\mathcal{U}, \mathcal{V}$). Finally, symbol $\triangleq$ is used in definition statements.
**Definitions.** Let $\mathcal{U} = \{u_1, \dots, u_n\}$ be a set of *users* and $\mathcal{V} = \{v_1, \dots, v_m\}$ be a set of *items*. Let $\mathcal{R}$ be a set of tuples $t_{ij} = (u_i, v_j, r_{ij})$, where $r_{ij}$ is a nonnegative number referred to as the *rating* given by user $u_i$ to the item $v_j$, and let $\mathbf{R} \in \Re^{n\times m}$ be a matrix whose $ij^{th}$ element contains the rating $r_{ij}$ if the tuple $t_{ij}$ belongs in $\mathcal{R}$, and zero otherwise.

### 2.1 From PureSVD to EigenRec

A recent successful example of latent-factor-based top-N recommendation algorithm is *PureSVD* [11]. This algorithm considers all missing values in the user-item ratings matrix, $\mathbf{R}$, as zeros, and produces recommendations by estimating $\mathbf{R}$ by the factorization

$$\hat{\mathbf{R}} = \mathbf{U_f}\mathbf{\Sigma_f}\mathbf{Q_f}^\intercal, \tag{1}$$

where $\mathbf{U_f}$ is an $n \times f$ orthonormal matrix, $\mathbf{Q_f}$ is an $m \times f$ orthonormal matrix, and $\mathbf{\Sigma_f}$ is an $f \times f$ diagonal matrix containing the $f$ largest singular values. The rows of matrix $\hat{\mathbf{R}}$ contain the recommendation vectors for every user in the system.

This matrix can be expressed in a different form that can provide more insight into the way PureSVD works, making it at the same time more amenable to generalizations. In particular, consider the full singular value decomposition of the ratings matrix $\mathbf{R}$:

$$\mathbf{R} = \mathbf{U}\mathbf{\Sigma}\mathbf{Q}^\intercal. \tag{2}$$

If we multiply equation (2) from the right with the orthonormal matrix $\mathbf{Q}$, we get

$$\mathbf{R}\mathbf{Q} = \mathbf{U}\mathbf{\Sigma}. \tag{3}$$

Now if we use $\mathbf{I_f}$ to denote the $f \times f$ identity matrix and we multiply again from the right with the $m \times m$ matrix $\begin{pmatrix} \mathbf{I_f} & \mathbf{0} \\ \mathbf{0} & \mathbf{0} \end{pmatrix}$, we get

$$\begin{aligned} \mathbf{R}\begin{pmatrix} \mathbf{Q_f} & \mathbf{0} \end{pmatrix} &= \mathbf{U}\begin{pmatrix} \mathbf{\Sigma_f} & \mathbf{0} \\ \mathbf{0} & \mathbf{0} \end{pmatrix} \\ \Rightarrow \mathbf{R}\mathbf{Q_f} &= \mathbf{U_f}\mathbf{\Sigma_f}. \end{aligned} \tag{4}$$

Substituting equation (4) in (1) gives

$$\hat{\mathbf{R}} = \mathbf{R}\mathbf{Q_f}\mathbf{Q_f}^\intercal. \tag{5}$$

Therefore the recommendation matrix of PureSVD can be expressed only in terms of the ratings matrix and matrix $\mathbf{Q_f}$. To get an intuitive understanding of (5) – and thereby to clarify the way PureSVD produces recommendations – it is worthwhile to give a small example.

Suppose user $u_i$ has rated only three items, namely item $v_1, v_3$ and $v_5$, with ratings $r_{i1} = 1, r_{i3} = 5$, and $r_{i5} = 3$, respectively. Also, assume for simplicity that we only need to decide whether we should recommend to this user, item 2 or item 4. What is PureSVD's solution to this dilemma? To answer this question let us follow the underlying computation for our example user, based on (5):

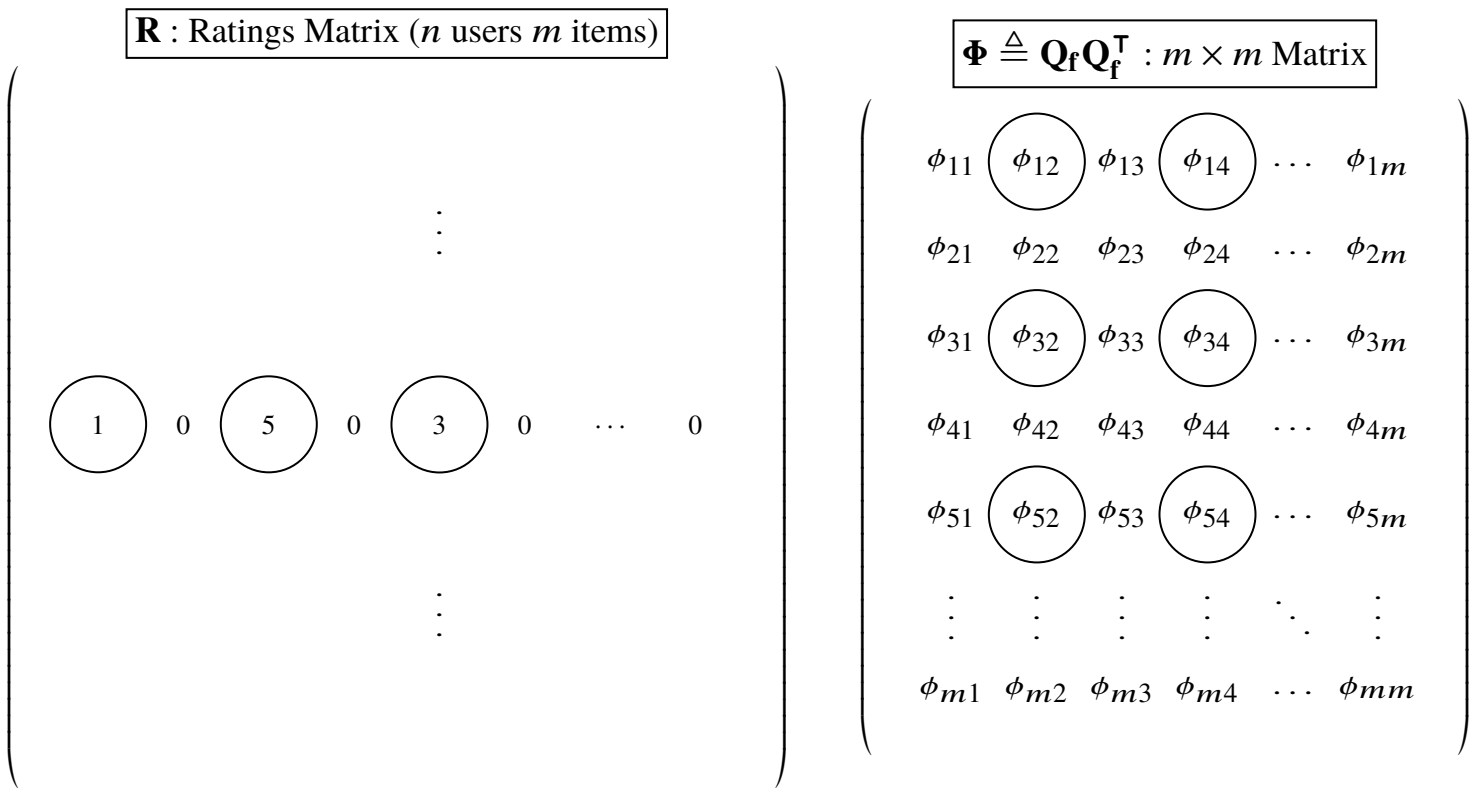

It is clear that PureSVD's scores[4] for items 2 and 4 will be

$$\begin{aligned} \pi_{i2} &= 1 \times \phi_{12} + 5 \times \phi_{32} + 3 \times \phi_{52} \\ \pi_{i4} &= 1 \times \phi_{14} + 5 \times \phi_{34} + 3 \times \phi_{54}. \end{aligned}$$

Notice that $\pi_{i2}$ and $\pi_{i4}$ are expressed only in terms of the known ratings of user $u_i$ and the elements of an $m \times m$ symmetric matrix which we denote for simplicity $\mathbf{\Phi}$. The $ij^{th}$ element of $\mathbf{\Phi}$ relates items $v_i$ and $v_j$. From a recommendation point of view, we can see that PureSVD treats these elements as measures of "closeness" or "similarity" between the corresponding items. For example, if $\phi_{12}, \phi_{52}, \phi_{14}, \phi_{54}$ in the above expression had the exact same value, PureSVD would recommend e.g. item 2 over item 4, only if item 2 was "more related" to item 3 than to item 4; i.e. if $\phi_{32}$ was larger than $\phi_{34}$. Therefore, informally, we can see that PureSVD's recommendation rule can be summed-up to the following:

> *Recommend to each user, the items that are more similar (in the $\mathbf{Q_f Q_f^\intercal}$ sense) to the items she has already seen.*

From the above discussion it becomes clear that PureSVD's performance is tied to the implicit choice of matrix $\mathbf{Q_f}$. However, from the definition of the singular value decomposition we know that $\mathbf{Q_f}$ contains the orthonormal set of eigenvectors that correspond to the $f$ principal eigenvalues of the symmetric matrix $\mathbf{R}^\intercal\mathbf{R}$ – and the elements of this matrix have a very intuitive interpretation in recommender systems parlance. In particular,

$$\mathbf{R}^{\intercal}\mathbf{R} = \mathit{items}\overset{\textit{users}}{\begin{bmatrix} — & \mathbf{r_i}^{\intercal} & — \end{bmatrix}} \times \mathit{users}\overset{\textit{items}}{\begin{bmatrix} | \\ \mathbf{r_j} \\ | \end{bmatrix}}$$

$$= \mathit{items}\overset{\textit{items}}{\begin{bmatrix} & \boxed{\cdot} & \end{bmatrix}} \quad \underbrace{\|\mathbf{r_i}\|\|\mathbf{r_j}\|}_{\textit{scaling}} \cdot \underbrace{\cos\theta_{ij}}_{\textit{similarity}} .$$

[4] Remember that these "scores" are by definition the elements that replace the previously zero-valued entries of the original ratings matrix $\mathbf{R}$, after its reconstruction using only the $f$ largest singular dimensions.

Thus, the $ij^{th}$ element of $\mathbf{R}^{\intercal}\mathbf{R}$ can be interpreted as the traditional cosine-based inter-item similarity score, scaled up by a factor related to the popularity of the items $v_i, v_j$ as expressed in the ratings matrix. Therefore, we see that the latent factor model of PureSVD is essentially built from the eigendecomposition of a *scaled cosine-based inter-item similarity matrix*.

From a purely computational perspective, this observation reduces the extraction of PureSVD's recommendation matrix to the calculation of the $f$ principal eigenvectors of an $m \times m$ symmetric matrix; a fact that can decrease markedly its overall computational and storage needs. More importantly, from a modeling perspective, the above observation places PureSVD in the center of a family of latent factor methods that can be readily obtained using inter-item proximity matrices that allow for (a) *different similarity functions* and (b) *different scaling functions*. We denote this family EigenRec, and we will show that even the simplest of its members can lead to high-quality results in challenging recommendation scenarios (Section 4).

## 2.2 EigenRec model definitions

Building on the above discussion, in this section we define formally the components of the EigenRec framework.

**Inter-Item Proximity Matrix A**. The Inter-Item Proximity matrix is designed to quantify the relations between the elements of the item space, as properly scaled pure similarity scores. Specifically, matrix $\mathbf{A} \in \Re^{m\times m}$ is a symmetric matrix, with its $ij^{th}$ element given by:

$$A_{ij} \triangleq \xi(i,j) \cdot \kappa(i,j), \tag{6}$$

where $\xi(\cdot,\cdot) : \mathcal{V} \times \mathcal{V} \mapsto [0,\infty)$ is a symmetric scaling function and $\kappa(\cdot,\cdot) : \mathcal{V} \times \mathcal{V} \mapsto \Re$ is a symmetric similarity function.

***Scaling Component.*** The definition of the scaling function can be done in many different ways, subject to various aspects of the recommendation problem at hand. In this work, we use this function as an easy way to regulate how much the inter-item proximity scores are affected by the prior popularity of the corresponding items. This was found to be very important for the overall recommendation quality as we will see in the experimental section of our paper. In particular, for the scaling function $\xi(\cdot,\cdot)$, we use the simple symmetric function

$$\xi(i,j) \triangleq (\|\mathbf{r_i}\|\|\mathbf{r_j}\|)^d. \tag{7}$$

where $\mathbf{r_i}$ denotes the $i^{th}$ column of matrix $\mathbf{R}$. Notice that the definition of the scaling function allows the final inter-item proximity matrix to be written in factorial form:

$$\mathbf{A} = \mathbf{SKS} \tag{8}$$

where

$$\mathbf{S} \equiv \mathbf{S}(d) \triangleq \text{diag}\{\|\mathbf{r_1}\|, \|\mathbf{r_2}\|, \dots, \|\mathbf{r_m}\|\}^d \tag{9}$$

and where matrix $\mathbf{K}$ (the $ij^{th}$ element of which is defined to be $\kappa(i, j)$), denotes the pure similarity component.

***Similarity Component.*** The definition of the similarity matrix $\mathbf{K}$ can be approached in several ways, depending on the nature of the recommendation task, the size of the itemset etc. Note that the final offline computational cost of the method may depend significantly on the choice of matrix $\mathbf{K}$ – especially when this matrix needs to be explicitly computed in advance or learned from the data. Having this in mind, in this work we propose using three widely used and simple similarity matrices that were found to be able to attain good results, while being easily manageable from a computational standpoint: (a) the *Cosine Similarity*, (c) the *Pearson-Correlation Similarity* and, (c) the *Jaccard Similarity*.

*Cosine Similarity* $\mathbf{K_{COS}}$. The similarity function $\kappa(\cdot, \cdot)$ is defined to be the cosine of the angle between the vector representation of the items $v_i, v_j$,

$$K_{ij} \triangleq \cos(v_i, v_j). \tag{10}$$

*Pearson Similarity* $\mathbf{K_{PC}}$. The similarity score between two items $v_i$ and $v_j$ is defined as the $ij^{th}$ element of matrix $\mathbf{K_{PC}}$ which is given by

$$K_{ij} \triangleq \frac{C_{ij}}{\sqrt{C_{ii}C_{jj}}}, \tag{11}$$

with $C_{ij}$ denoting the covariance between the vector representation of the items $v_i, v_j$.

*Jaccard Similarity* $\mathbf{K_{JAC}}$. The Jaccard similarity between two items is defined as the ratio of the number of users that have rated both items to the number of users that have rated at least one of them. Specifically,

$$K_{ij} \triangleq \frac{|\mathcal{R}_i \cap \mathcal{R}_j|}{|\mathcal{R}_i \cup \mathcal{R}_j|}, \tag{12}$$

where $\mathcal{R}_i$ the set of users that have rated item $i$.

**Recommendation Matrix $\mathbf{\Pi}$**. The final recommendation matrix contains the recommendation vectors for each user in the system. In particular, for each user $u_i$ the corresponding personalized recommendation vector is given by:

$$\boldsymbol{\pi}_\mathbf{i}^\intercal \triangleq \mathbf{r}_\mathbf{i}^\intercal \mathbf{V}\mathbf{V}^\intercal, \tag{13}$$

where $\mathbf{r}_\mathbf{i}^\intercal$ the ratings of user $u_i$ and $\mathbf{V} \in \Re^{m \times f}$ is the matrix whose columns contain the $f$ *principal orthonormal eigenvectors* of the inter-item proximity matrix $\mathbf{A}$. Observe that since $\mathbf{A}$ is real and symmetric, its eigenvectors are real and can be chosen to be orthogonal to each other and of unity norm.

**PureSVD within EigenRec.** Clearly, the final recommendation matrix of PureSVD coincides with that produced by EigenRec, using the similarity matrix $\mathbf{K_{cos}}$ and the standard scaling matrix $\mathbf{S}$ with parameter $d = 1$,

$$\text{PureSVD}(\mathbf{R}) \equiv \text{EigenRec}(\mathbf{K}_{\text{cos}}, \mathbf{S}(d = 1)). \tag{14}$$

Furthermore, a closer look at our derivations in §2.1 reveals that EigenRec with cosine similarity matrix and Euclidean norm scaling for a given parameter $d$, actually coincides with PureSVD applied to a *modified ratings matrix*. In particular,

$$\text{EigenRec}(\mathbf{K}_{\text{cos}}, \mathbf{S}(d)) \equiv \text{PureSVD}(\tilde{\mathbf{R}}) \tag{15}$$

where

$$\tilde{\mathbf{R}} \triangleq \mathbf{RD}, \qquad \mathbf{D} = \text{diag}\{\|\mathbf{r_1}\|, \|\mathbf{r_2}\|, \dots, \|\mathbf{r_m}\|\}^{d-1}. \tag{16}$$

Notice that since $\mathbf{D}$ is a diagonal matrix, from (16) it follows that to get the final matrix $\tilde{\mathbf{R}}$ we need to multiply each column $j$ of the original ratings matrix by $\|\mathbf{r_j}\|^{d-1}$. Now, for values of $d$ less than 1 which – as we will see in the experimental section of this paper – yield the best top-N recommendation performance, the above operation "penalizes" each item by a factor related to a measure of its prior popularity. Seeing PureSVD within our framework hints that its implicitly chosen value for the parameter $d$, makes it overly sensitive to the prior popularity of the items and, as we will see, it is exactly this suboptimal default choice of scaling that inevitably hinders its potential.

Having defined formally the components of our recommendation framework, we are now ready for our "computational interlude", where we discuss the details behind building the latent space efficiently.

## 3 Building the latent space

At the computational core of EigenRec is the extraction of the principal eigenvectors of a sparse symmetric linear operator. For modest sized problems this can be done easily in a single machine using mature eigensolvers written in high-performance compiled languages. Wrappers for calling these solvers are typically available in virtually every high-level programming language. One of our main goals in this work however, is to ensure the practical application of EigenRec even in the context of very large datasets. To this end, in this section we provide an overview of the computational aspects of EigenRec and discuss in detail its parallel implementation in distributed memory environments.

### 3.1 EigenRec computation: algorithm and parallel implementation

The specific properties of our model (symmetry and sparsity), allow us to use the symmetric *Lanczos algorithm* [26] – an iterative Krylov subspace method for the solution of large and sparse eigenvalue problems – to build the latent space, $\mathbf{V}$, and produce the recommendation lists efficiently. Given a matrix $\mathbf{A} \in \Re^{m \times m}$ and an initial unit vector $\mathbf{q}$, the corresponding Krylov subspace of size $\ell$ is given by $\mathcal{K}_\ell(\mathbf{A}, \mathbf{q}) \triangleq \text{span}\{\mathbf{q}, \mathbf{Aq}, \mathbf{A}^2\mathbf{q}, \dots, \mathbf{A}^{\ell-1}\mathbf{q}\}$. By forming an orthonormal basis for $\mathcal{K}_\ell$, Krylov subspace methods can be used to solve several types of numerical problems. In this section we describe the application of Lanczos algorithm to build our latent factor subspace, $\mathbf{V}$.

**Lanczos Algorithm.** The algorithm starts by choosing a random vector $\mathbf{q}$, and builds an orthonormal basis $\mathbf{Q_j}$ of the Krylov subspace $\mathcal{K}_j(\mathbf{A}, \mathbf{q})$, one column at a time. In this orthonormal basis $\mathbf{Q_j}$, the operator $\mathbf{A}$ is represented by a real symmetric tridiagonal matrix,

$$\mathbf{T_j} = \begin{bmatrix} \alpha_1 & \beta_1 & & \\ \beta_1 & \alpha_2 & \ddots & \\ & \ddots & \ddots & \beta_{j-1} \\ & & \beta_{j-1} & \alpha_j \end{bmatrix}, \tag{17}$$

which is also built up one row and column at a time [4], using the recurrence,

$$\mathbf{AQ_j} = \mathbf{Q_j T_j} + \mathbf{r e_j^{\intercal}} \quad \text{with} \quad \mathbf{Q_j^{\intercal} r} = 0. \tag{18}$$

In exact arithmetic, the orthonormality of the Krylov subspace is preserved implicitly by the three-term recurrence in (18). However, in most real-case applications of Lanczos, orthogonality of the Krylov subspace is maintained explicitly. The leading eigenvectors of $\mathbf{A}$ can be approximated by first computing (at any step $j$) the eigendecomposition of $\mathbf{T_j}$,

$$\mathbf{T_j} = \mathbf{\Xi \Theta \Xi^{\intercal}}, \tag{19}$$

and then forming the Ritz vectors $\mathbf{Q_j}\boldsymbol{\xi}_i$, $i = 1, \ldots, j$. The eigenvalues of $\mathbf{T_j}$ (Ritz values) approximate those of $\mathbf{A}$, with the ones located at the periphery of the spectrum being approximated first. In practice, the latter implies fast convergence of Lanczos towards invariant subspaces associated with the leading eigenvalues.

To measure the error of the approximation of each latent factor, we need to compute the residual norm of each approximate eigenpair. It can be shown that the residual norm of the $i^{th}$ approximate eigenpair at the $j^{th}$ Lanczos step satisfies the equation, $\delta_i^{(j)} = |\beta_j \Xi_{ji}|, i = 1, \ldots, j$, and thus it suffices to monitor only the subdiagonal element $\beta_j$ of $\mathbf{T}$ and the last row of $\mathbf{\Xi}$ [4]. The algorithm for the computation of $\mathbf{V}$ and the final recommendation matrix $\mathbf{\Pi}$ for the whole set of users is given in Algorithm 1.

**Algorithm 1** EigenRec

**Input:** Inter-Item proximity matrix $\mathbf{A} \in \Re^{m \times m}$. Ratings Matrix $\mathbf{R} \in \Re^{n \times m}$. Latent Factors $f$.
**Output:** Matrix $\mathbf{\Pi} \in \Re^{n \times m}$ whose rows are the recommendation vectors for every user.

$\mathbf{q_0} = 0$, set $\mathbf{r} \leftarrow \mathbf{q}$ as a random vector
$\beta_0 \leftarrow \|\mathbf{r}\|_2$
**for** $j \leftarrow 1, 2, \ldots,$ **do**
  $\mathbf{q_j} \leftarrow \mathbf{r}/\beta_{j-1}$
  $\mathbf{r} \leftarrow \mathbf{A q_j}$
  $\mathbf{r} \leftarrow \mathbf{r} - \mathbf{q_{j-1}}\beta_{j-1}$
  $\alpha_j \leftarrow \mathbf{q_j^{\intercal} r}$
  $\mathbf{r} \leftarrow \mathbf{r} - \mathbf{q_j}\alpha_j$
  $\mathbf{r} \leftarrow (\mathbf{I} - \mathbf{Q_j Q_j^{\intercal}})\mathbf{r},$ ▷ where $\mathbf{Q_j} = [\mathbf{q_1}, \ldots, \mathbf{q_j}]$
  $\beta_j \leftarrow \|\mathbf{r}\|_2$
  Solve the tridiagonal problem $\mathbf{T_j \Xi_j} = \mathbf{\Theta_j \Xi_j}$
  Form the $j$ approximate eigenvectors $\mathbf{Q_j \Xi_j}$ of $\mathbf{A}$
  If the $f$ top eigenvectors have converged, stop.
**end for**
Collect the $f$ converged latent factors in a matrix $\mathbf{V}$.
**return** $\mathbf{\Pi} \leftarrow \mathbf{RVV^{\intercal}}$

**Computational Cost.** In terms of computational complexity, the most expensive operations of Lanczos are the MV product in Step 5, and the reorthogonalization procedure in Step 9. The total cost introduced by the Matrix×Vector (MV) products in $j$ Lanczos steps amounts to $O(j \cdot nnz)$, with $nnz$ denoting the number of non-zero entries. At the same time, making the $j^{th}$ Lanczos vector orthogonal to the previous $j-1$ ones requires $O(jm)$ floating point operations. The latter implies that as $j$ increases, reorthogonalization costs will eventually become the main bottleneck. On the other hand, the memory complexity is linear to the number of Lanczos steps and it is essentially dominated by the need to store all vectors produced by Lanczos.

**Parallel Implementation.** While Lanczos is an inherently serial procedure – in the sense that the next iteration starts only after the previous one is completed – we can speed-up its application by performing its computations in parallel. More specifically, let us assume a distributed memory environment with $P$ processors. For simplicity, we discuss the parallel implementation for the latent space construction of an inter-item proximity matrix $\mathbf{A}$ that can be written in a simple product form, $\mathbf{A} = \mathbf{W}^\intercal\mathbf{W}$, as in the case e.g. of the Cosine similarity matrix[5]. Since typically the ratings are concentrated in small regions of the overall rating matrices, in order to achieve better load balancing among the processors we distribute matrix $\mathbf{W}^\intercal$ across all $P$ processors based on the number of non-zero ($nnz$) entries; i.e. different processors are assigned a different number of rows, so that all processors share roughly the same number of non-zero entries.

*The MV product:* The MV product between $\mathbf{A}$ and a vector $\mathbf{q}$ in Step 5 of Algorithm 1, can be achieved by a two-stage procedure where we first compute $\hat{\mathbf{q}} = (\mathbf{W}\mathbf{q})$ followed by $\mathbf{y} = \mathbf{W}^\intercal\hat{\mathbf{q}}$. Assuming that $\mathbf{W}^\intercal$ is distributed row-wise (thus $\mathbf{W}$ is distributed column-wise), the only need for communication appears when performing $\hat{\mathbf{q}} = (\mathbf{W}\mathbf{q})$ and consists of a simple `allreduce` operation to sum the local contribution of each process.

*The inner products:* The inner product is the second operation of Lanczos which demands communication among the processors. It is a two-stage procedure where in the first stage each processor computes its local part of the global inner product, while in the second stage the local inner products (a scalar value per processor) are summed by `allreduce` and the final value is distributed to all processors.

*Reorthogonalization:* Similarly to the above computations, the reorthogonalization step during the $j^{th}$ Lanczos iteration

$$\mathbf{q}'_{\mathbf{j+1}} = \mathbf{q}_{\mathbf{j+1}} - \mathbf{Q}_{\mathbf{j}}\mathbf{Q}_{\mathbf{j}}^\intercal\mathbf{q}_{\mathbf{j+1}},$$

is performed by a two-stage procedure where we first compute $\hat{\mathbf{y}} = \mathbf{Q}_{\mathbf{j}}^\intercal\mathbf{q}_{\mathbf{j+1}}$ followed by $\mathbf{q}'_{\mathbf{j+1}} = \mathbf{q}_{\mathbf{j+1}} - \mathbf{Q}_{\mathbf{j}}\hat{\mathbf{y}}$. The only need for communication among the different processors appears when performing $\hat{\mathbf{y}} = \mathbf{Q}_{\mathbf{j}}^\intercal\mathbf{q}_{\mathbf{j+1}}$ and is of the `allreduce` form.

*The vector updates:* The vector updates are trivially parallel.

Similar approaches can be followed if matrix $\mathbf{A}$ is given by more general expressions. In general, as it is the case with all Krylov subspace methods, Lanczos does not require matrix $\mathbf{A}$ be explicitly formed; only a routine that is able to perform the MV products with $\mathbf{A}$ is necessary. Furthermore, having the inter-item proximity matrix explicitly formed is not advised since any explicit formation will probably be much more dense; resulting in an unnecessary raise of the computational time spent on Lanczos compared to the same run using the product form. For a more extensive discussion on Lanczos, its different variants, as well as additional discussion on its parallelization strategies, we refer to [16].

[5] for which, if we assume scaling parameter $d$, matrix $\mathbf{W}$ equals $\mathbf{R}\,\mathrm{diag}\{\|\mathbf{r_1}\|, \|\mathbf{r_2}\|, \dots, \|\mathbf{r_m}\|\}^{d-1}$.

## 4 Experimental evaluation

### 4.1 Datasets and metrics

The recommendation quality of our method was tested utilizing data originated from two recommendation domains, namely **Movie Recommendation** – where we exploit the standard `MovieLens1M` and `MovieLens100K` datasets [19] that have been used widely for the qualitative evaluation of recommender systems; and **Song Recommendation** – where we used the `Yahoo!R2Music` dataset [41] which represents a snapshot of the Yahoo!Music community's preferences for different songs. More details about the datasets used can be found in [19,41].

#### *4.1.1 Metrics*

For our qualitative experiments, except for the standard **Recall** and **Precision** metrics [3,11], we also use a number of other well known *utility-based* ranking indices, that assume that the utility of a recommended item is discounted by a factor related to its position in the final recommendation list [34]. Depending on the decay of the positional discount down the list we have the:

**Normalized Discounted Cumulative Gain,** which assumes that the ranking positions are discounted *logarithmically fast* [5,34] and is defined by:

$$\mathrm{NDCG}@k = \frac{\mathrm{DCG}@k(\mathbf{y},\boldsymbol{\pi})}{\mathrm{DCG}@k(\mathbf{y},\boldsymbol{\pi}^\star)}, \tag{20}$$

with

$$\mathrm{DCG}@k(\mathbf{y},\boldsymbol{\pi}) = \sum_{q=1}^{k} \frac{2^{y_{\pi_q}} - 1}{\log_2(2+q)}, \tag{21}$$

where $\mathbf{y}$ is a vector of the relevance values for a sequence of items, $\pi_q$ is the index of the $q^{th}$ item in the recommendation list $\boldsymbol{\pi}$, and $\boldsymbol{\pi}^\star$ is the optimal ranking of the items with respect to the relevant scores (see [5] for details).

**RScore,** which assumes that the value of recommendations declines *exponentially fast* to yield the following score:

$$\mathrm{R}(\alpha) = \sum_{q} \frac{\max(y_{\pi_q} - d, 0)}{2^{\frac{q-1}{\alpha-1}}}, \tag{22}$$

where $\alpha$, controls the exponential decline and is referred to as the half-life parameter (see [34] for details).

**Mean Reciprocal Rank,** which assumes a slower decay than R-Score but faster than NDCG. MRR is the average of the reciprocal rank scores of the users, defined as follows:

$$\mathrm{RR} = \frac{1}{\min_q\{q : y_{\pi_q} > 0\}}. \tag{23}$$

### 4.2 Top-N recommendation quality

We compare EigenRec against a number of methods of the graph-based top-$N$ recommendation family, that are considered to be promising in dealing with sparsity [12]. Generally speaking, graph-based recommendation methods represent the recommender database as a bipartite user-item graph, $\mathcal{G} = \{\{\mathcal{V}, \mathcal{U}\}, \mathcal{E}\}$ where $\mathcal{E} = \{e_{ij} \mid i \in \mathcal{V}, j \in \mathcal{U}$ such that $t_{ij} \in \mathcal{R}\}$, and try to estimate similarity or distance measures between the nodes which can be used for the computation of ranked lists of the items with respect to each user.

The five competing methods used in our experiments are: the *Pseudo-Inverse of the user-item graph Laplacian* (L†), the *Matrix Forest Algorithm* (MFA), the *Regularized Commute Time* (RCT), the *Markov Diffusion Kernel* (MD) and the *Relative Entropy Diffusion* (RED). Below we give their formal definitions.

**The pseudoinverse of the Laplacian.** This matrix contains the inner products of the node vectors in a Euclidean space where the nodes are exactly separated by the commute time distance [15]. For the computation of the $\mathbf{G}_{\mathbf{L}^{\dagger}}$ matrix we used the formula:

$$\mathbf{G}_{\mathbf{L}^{\dagger}} \triangleq (\mathbf{L} - \frac{1}{n+m}\mathbf{e}\mathbf{e}^{\intercal})^{-1} + \frac{1}{n+m}\mathbf{e}\mathbf{e}^{\intercal}, \quad (24)$$

where $\mathbf{L}$ is the Laplacian of the graph model of the recommender system, $n$, the number of users, and $m$, the number of items (see [14] for details).

**The MFA matrix.** MFA matrix contains elements that also provide similarity measures between nodes of the graph by integrating indirect paths, based on the matrix-forest theorem [9]. Matrix $\mathbf{G}_{\mathbf{MFA}}$ was computed by

$$\mathbf{G}_{\mathbf{MFA}} \triangleq (\mathbf{I} + \mathbf{L})^{-1}, \quad (25)$$

where $\mathbf{I}$, the identity matrix and $\mathbf{L}$, defined above.

**Markov Diffusion Kernel.** As discussed in [14] the underlying hypothesis behind this kernel is that similar nodes diffuse in a similar way through the graph. Concretely, if we define a stochastic matrix $\mathbf{P} \triangleq \mathbf{D}^{-1}\mathbf{A}$, where $\mathbf{A}$ is the adjacency matrix of the graph and $\mathbf{D}$, a diagonal matrix containing the outdegrees of the graph nodes, the Markov diffusion kernel with parameter $t$ is defined by

$$\mathbf{G}_{\mathbf{MD}} \triangleq \mathbf{Z}(t)\mathbf{Z}^{\intercal}(t), \quad \text{with} \quad \mathbf{Z}(t) \triangleq \frac{1}{t}\sum_{\tau=1}^{t}\mathbf{P}^{\tau}. \quad (26)$$

Extensive experiments done by the authors in [14] suggest that the Markov diffusion kernel does particularly well in collaborative recommendation tasks.

**Relative Entropy Diffusion Matrix.** This similarity matrix is based on the Kullback-Leibler divergence between distributions and it is defined by

$$\mathbf{G}_{\mathbf{RED}} \triangleq \mathbf{Z}(t)\log(\mathbf{Z}^{\intercal}(t)) + \log(\mathbf{Z}(t))\mathbf{Z}^{\intercal}(t), \quad (27)$$

where $\mathbf{Z}(t)$ is defined as previous. As with the Markov diffusion kernel, $t$ is a parameter of the model.

**Regularized Commute Time Kernel.** Finally, the Regularized Commute Time is defined by

$$\mathbf{G}_{\mathbf{RCT}} \triangleq (\mathbf{D} - \alpha\mathbf{A})^{-1}, \quad (28)$$

and its $ij^{th}$ element denotes the discounted cumulated probability of visiting node $j$ when starting from node $i$ [14,43].

For our experiments we tested each method for many different values of the parameters for every dataset and we report the best results achieved for each experiment. Table 1 shows the parametric range tested for each method. For further details about the competing methods the reader should see [9, 14] and the references therein.

**Table 1** Parameter selection for the competing recommendation methods

| Method | Parameters | Range Tested |
|---:|:---:|---:|
| PseudoInverse of the Laplacian | – | – |
| Matrix Forest Algorithm | – | – |
| Markov Diffusion Kernel | $t$ | $1, 2, \ldots, 10, 50, 100$ |
| Relative Entropy Diffusion Matrix | $t$ | $1, 2, \ldots, 10, 50, 100$ |
| Regularized Commute Time Kernel | $\alpha$ | $10^{-6}, 10^{-5}, \ldots, 0.99$ |

For our recommendation quality comparison tests we used the complete `MovieLens1M` dataset (denoted `ML1M`) and – following the dataset preparation approach used by Karypis *et al.* in [22] – a randomly selected subset of the Yahoo! Research Alliance Webscope Dataset (denoted `Yahoo`) with 3312 items and 7307 users.

Except for the *Standard Recommendation*, we also test the performance of our method in dealing with two very challenging and realistic scenarios that are linked to the inherent sparsity of typical recommender systems datasets. Namely, the *Long-Tail Recommendation*, where we evaluate the ability of our method in making useful recommendations of unpopular items, and the *Cold-Start Recommendation*, where we evaluate how well it does in recommending items for *New Users* in an existing recommender system (localized sparsity) as well as making recommendations for a *New Community* of users in the starting stages of the system.

### *4.2.1 Standard recommendations*

To evaluate the quality of EigenRec in suggesting top-$N$ items, we have adopted the methodology proposed by Cremonesi *et al.* in [11]. In particular, we form a probe set $\mathcal{P}$ by randomly sampling 1.4% of the ratings of the dataset, and we use each item $v_j$, rated with 5-star by user $u_i$ in $\mathcal{P}$ to create the test set $\mathcal{T}$. For each item in $\mathcal{T}$, we select randomly another 1000 unrated items of the same user, we rank the complete lists (containing 1001 items) using each of the competing methods, and we measure the respective recommendation quality.

First we test the recommendation performance of EigenRec in the MRR metric for scaling parameters in the range $[-2, 2]$ using all three similarity matrices. We choose the MRR metric for this test simply because it can summarize the recommendation performance in a single number which allows direct comparisons between different similarity matrices as well as different scaling parameters for each given matrix. Figure 1 reports the MRR scores as a function of the parameter $d$ for every case, using the number of latent factors that produces the best possible performance for each matrix.

We see that the best performance is achieved for small positive values of parameter $d$. This was true for every similarity matrix tested, and for both datasets. Notice that this parameter was included in our model as a means to control the sensitivity of the inter-item proximity scores to the prior popularity of the items under consideration. Our results suggest, that while this popularity is important (i.e. every time the best performing scaling factor was strictly positive), its contribution to the final matrix **A** should be weighted carefully so as not to overshadow the pure similarity component.

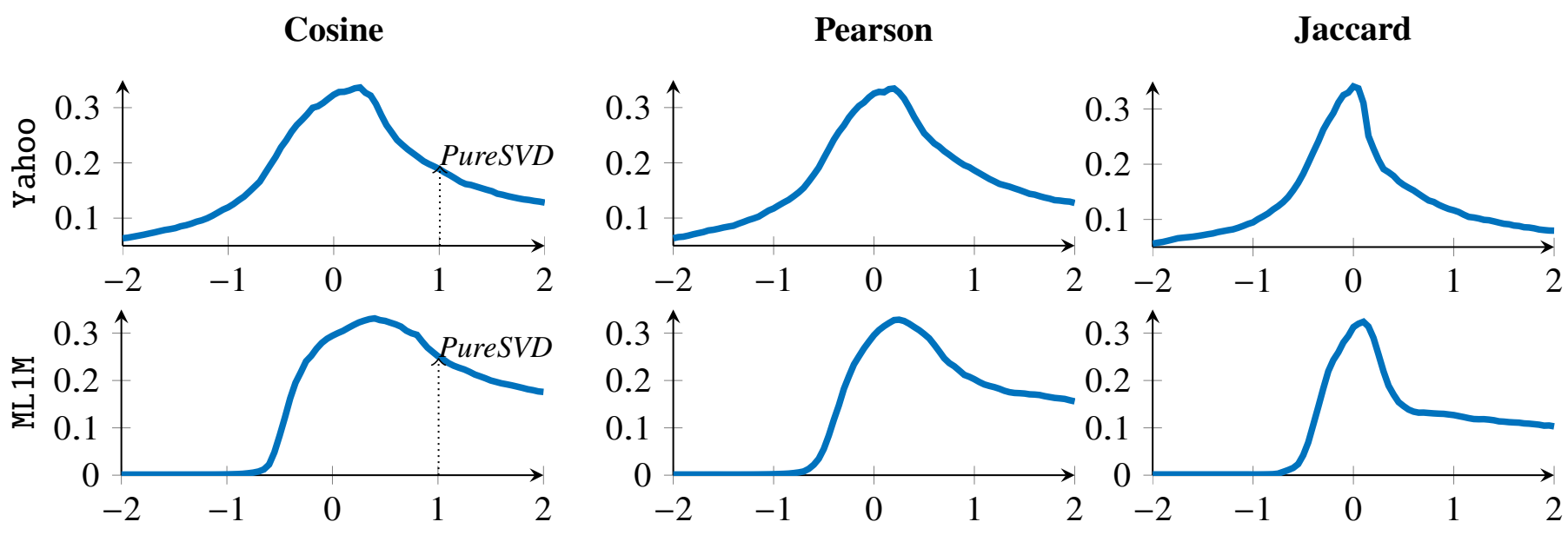

**Fig. 1** Recommendation performance of EigenRec on the MRR metric for scaling factors in the range [−2, 2] using all three similarity matrices

We see that all variations of our method outperform PureSVD every time, with the performance gap being significantly larger for the Yahoo dataset, which had a steeper performance decay as the scaling factors moved towards 1 (see Figure 1). Recall that the “black box” approach of the traditional PureSVD assumes cosine similarity (which is usually great) with scaling parameter $d$ equal to 1 (which is usually not). As can be seen in Figure 1, simply controlling parameter $d$ alone results to significant recommendation performance gains with respect to PureSVD. We find this particularly interesting, as it uncovers a fundamental limitation of the traditional PureSVD approach, that can be trivially alleviated with our approach.

We also compare EigenRec against the five graph-based methods mentioned in the beginning of this section. For these comparisons, we used the Jaccard similarity matrix. We tested each method for many different values of the parameters for every dataset and we report the best results achieved for each experiment. Figure 2 reports the Recall as a function of $N$ (i.e. the number of items recommended) the Precision as a function of the Recall, the Normalized Discounted Cumulative Gain as a function of $N$ and the RScore as a function of the halflife parameter $\alpha$, for the `Yahoo` (first row) and the `MovieLens1M` (second row) datasets. As for Recall($N$) and NDCG@N, we consider values of $N$ in the range $[1, \ldots, 20]$; larger values can be safely ignored for a typical top-$N$ recommendation task [11]. As we can see, EigenRec outperforms every other method considered, for all datasets and in all metrics, reaching for example, at $N = 10$ a recall around 60%. This means that 60% of the 5-starred items were presented in the top-10 out of the 1001 places in the recommendation lists of the respective users.

### *4.2.2 Long-Tail recommendations*

The distribution of rated items in recommender systems is long-tailed, i.e. most of the ratings are concentrated in a few very popular items, leaving the rest of the itemspace unevenly sparse. Of course, recommending popular items is an easy task, adding little utility in recommender systems; on the other hand, the task of recommending long-tail items adds *novelty* and *serendipity* to the users [11], and it is also known to increase substantially the profits of e-commerce companies [1,42]. The innate sparsity of the problem however – which is aggravated even more for long-tail items – presents a major challenge for the majority of state-of-the-art collaborative filtering methods.

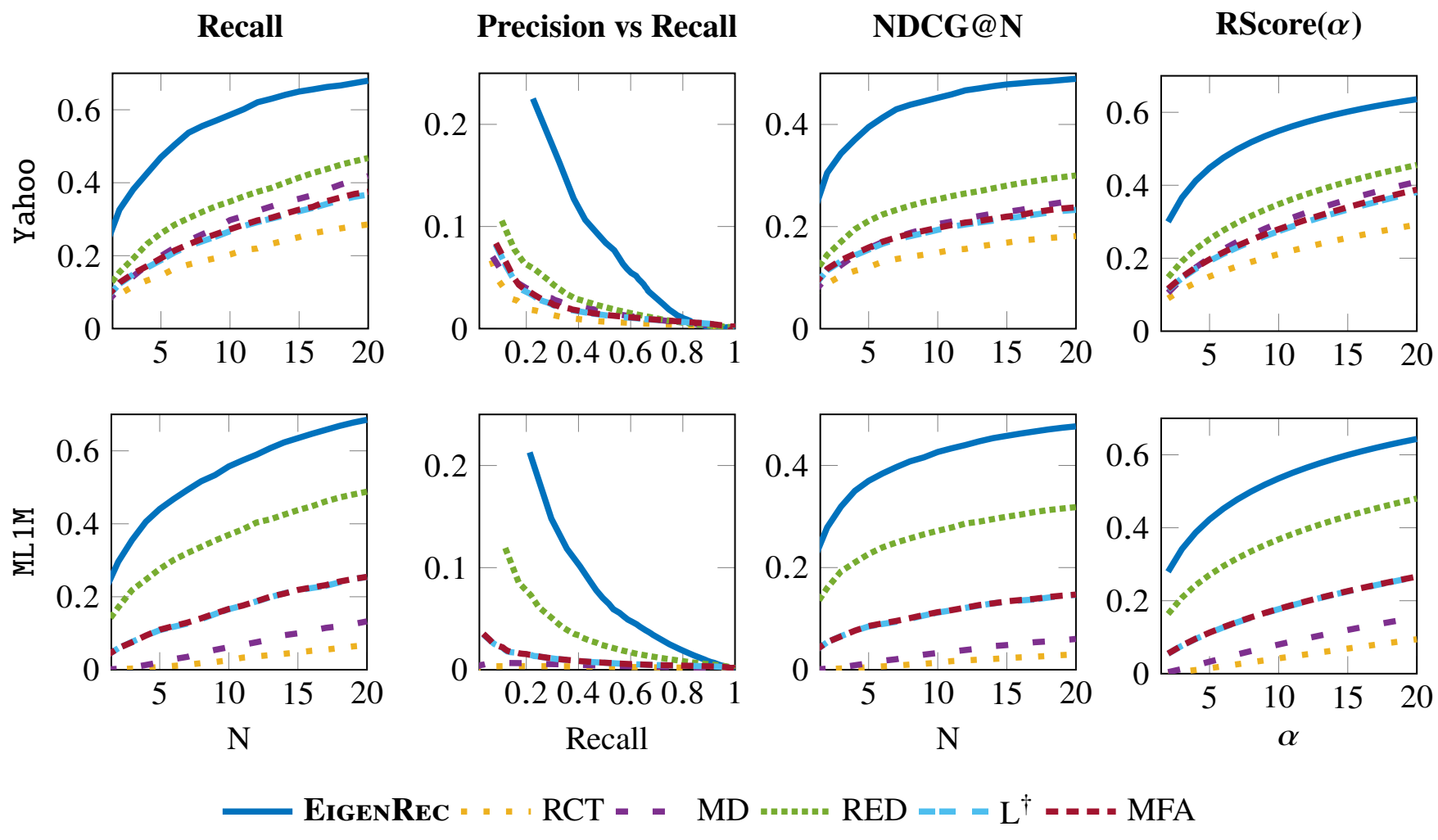

**Fig. 2** Evaluation of the recommendation quality using the Recall@N, Precision, NDCG@N and RScore metrics

To evaluate EigenRec in recommending long-tail items, we adopt the methodology described in [11]. In particular, we order the items according to their popularity which was measured in terms of number of ratings, and we partition the test set $\mathcal{T}$ into two subsets, $\mathcal{T}_{\text{tail}}$ and $\mathcal{T}_{\text{head}}$, that involve items originated from the long-tail, and the short-head of the distribution respectively. We discard the items in $\mathcal{T}_{\text{head}}$ and we evaluate EigenRec and the other algorithms on the $\mathcal{T}_{\text{tail}}$ test set, using the procedure explained in §4.2.1.

Having evaluated the performance of EigenRec in the MRR metric for all three similarity matrices, we obtained very good results for every case, with marginally better recommendation quality achieved for the Jaccard similarity component with 241 and 270 latent factors and scaling factor 0.2 and 0.4 for the `Yahoo` and the `MovieLens1M` datasets respectively. Proceeding with these parameter settings we run EigenRec against the other graph-based algorithms and we report the results in Figure 3. It is interesting to notice that MFA and $L^{\dagger}$ do particularly well in the long-tail recommendation task, especially in the sparser `Yahoo` dataset. They even manage to surpass RED, which had reached the second place when the popular items were included (Figure 2). Once again, we see that EigenRec achieves the best results, in all metrics and for both datasets.

We have seen that both in standard and long-tail recommendation scenarios, our approach gives very good results, consistently outperforming – besides PureSVD – a number of elaborate graph-based methods, known to work very well in uncovering nontrivial similarities through the exploitation of transitive relations that the graph representation of the data brings to light [12]. In our final set of experiments, presented next, we test the performance of EigenRec in dealing with sparsity in its most extreme manifestations; the Cold-Start Problems.

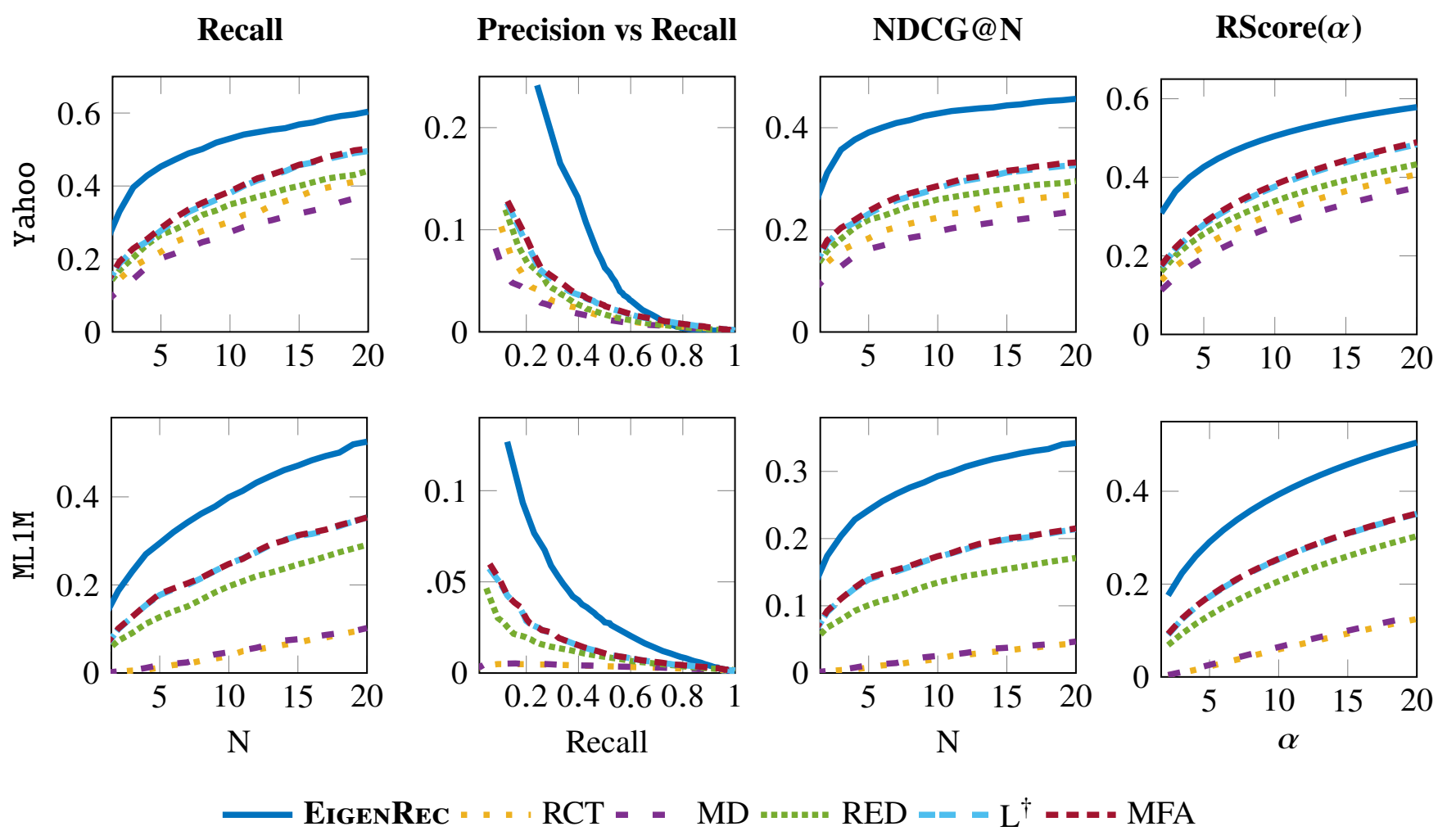

**Fig. 3** Evaluation of the Long-Tail recommendation quality using the Recall@N, Precision, NDCG@N and RScore metrics

### *4.2.3 Cold-Start recommendations*

The cold-start problem refers to the difficulty of making reliable recommendations due to an initial lack of ratings [8]. This is a very common problem faced by real recommender systems in their beginning stages, when the number of ratings for the collaborative filtering algorithms to uncover similarities between items or users are insufficient (*New Community Problem*). The problem can arise also when introducing new users to an existing system (*New Users Problem*); typically new users start with only a few ratings, making it difficult for the collaborative filtering algorithm to produce reliable personalized recommendations. This can be seen as a type of localized sparsity problem and it represents one of the ongoing challenges faced by recommender systems in operation.

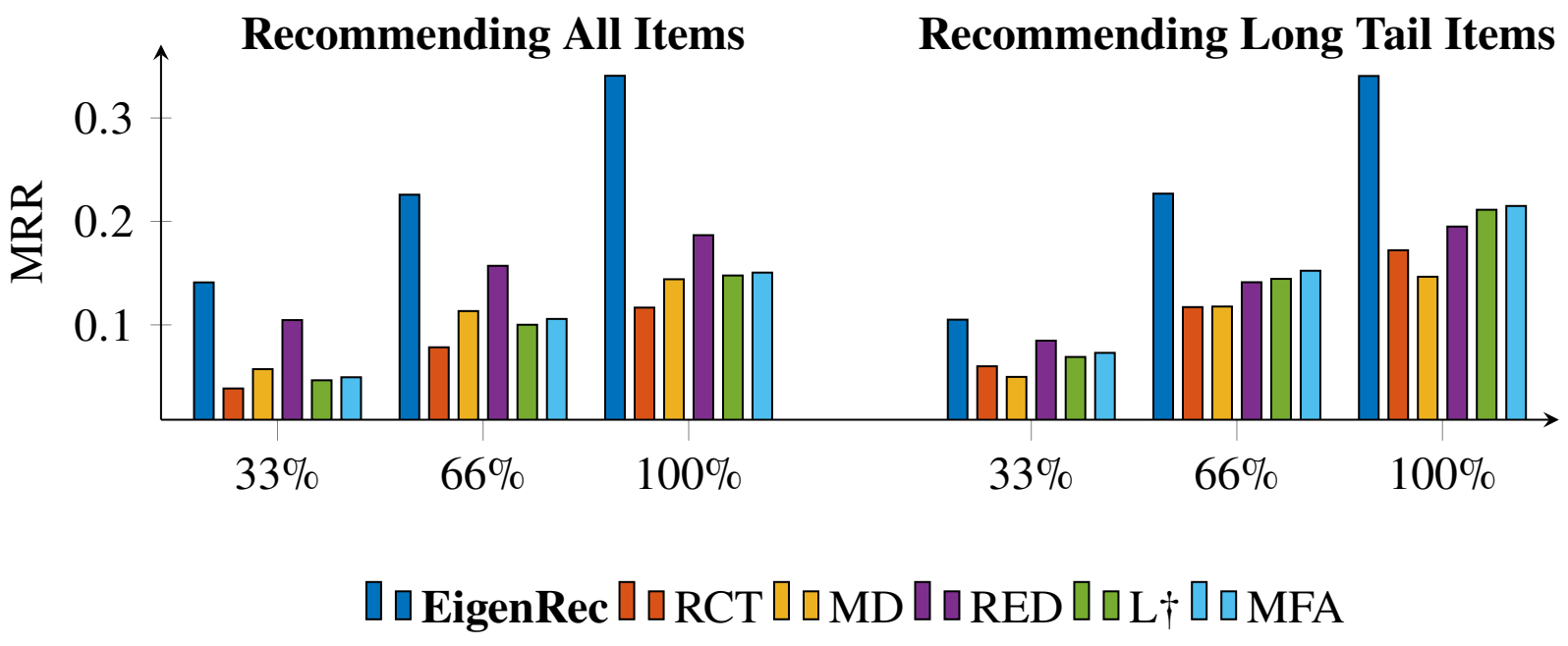

**Fig. 4** New-Community recommendation quality using the MRR metric

*New Community Problem:* To test the performance of EigenRec in dealing with the new community problem, we conduct the following experiment: We simulate the phenomenon by randomly selecting to include 33%, and 66% of the `Yahoo` dataset on two new artificially sparsified versions in such a way that the first dataset is a subset of the second. The idea is that these new datasets represent snapshots of the initial stages of the recommender system, when the community of users was new and the system was lacking ratings [27]. Then, we take the new community datasets and we create test sets following the methodology described in Section 4.2.1; we run all the algorithms and we evaluate their performance using the MRR, which makes it easier to compare the top-$N$ quality for the different stages in the system's evolution. We test for both standard and long-tail recommendations and we report the results in Figure 4. We clearly see that EigenRec outperforms every other algorithm, even in the extremely sparse initial stage where the system is lacking 2/3 of its ratings. In the figure, we report the qualitative results using the Cosine similarity this time, however, the performance of the three similarity components we propose was found to be equally good.

*New Users Problem:* In order to evaluate the performance of our algorithm in dealing with new users, we again use the `Yahoo` dataset and we run the following experiment. We randomly select 50 users having rated 100 items or more, and we randomly delete 95% of their ratings. The idea is that the modified data represent an "earlier version" of the dataset, when these users were new to the system, and as such, had fewer ratings. Then, we take the subset of the dataset corresponding to these new users and we create the test set as before, using 10% as a cut-off for the Probe Set this time, in order to have enough 5-rated movies in the Test Set to estimate reliably the performance quality. The results are presented in Figure 5. We see that EigenRec manages to outperform all competing algorithms in all metrics as before.

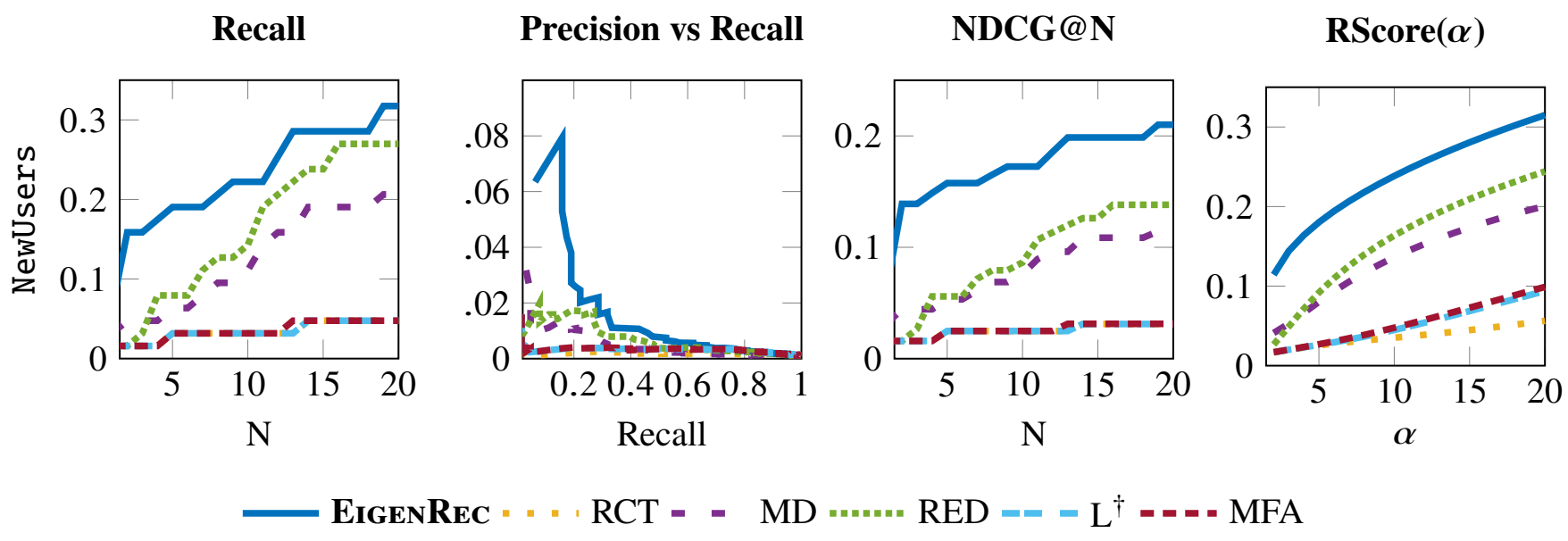

**Fig. 5** New-Users recommendation quality using the Recall@N, Precision, NDCG@N and RScore metrics

### 4.2.4 Discussion

The qualitative results presented above indicate that our method is able to produce high quality recommendations, alleviating significant problems related to sparsity. Let us mention here that the competing algorithms are considered among the most promising methods in the literature to address sparsity problems [12]. This was verified in our experiments as well. Indeed, our results clearly show that the graph-based methods perform very well with their comparative performance increasing with the sparsity of the underlying dataset, and reaching its maximum in the cold-start scenarios. EigenRec nonetheless managed to perform even better, in every recommendation setting considered, being at the same time by far the most economical method from a computational point of view. Note here that all competing methods require handling a graph of $m + n$ nodes (where $m$ the number of items and $n$ the number of users), with the extraction of the recommendation scores many times involving inversions of $(m + n)$-dimensional square matrices etc. – problems that easily become intractable as the population of users in the system increases. EigenRec, on the contrary having a significantly friendlier computational profile, denotes a qualitative and feasible option for realistic top-$N$ recommendation settings.

The choice of the scaling factor was found to be particularly significant for each and every pure similarity component. For the cosine similarity, in particular, we observed that the best results were always achieved for scaling parameters away from 1, making the traditional PureSVD algorithm, “qualitatively dominated” in every case considered. Regarding the best choice for the pure similarity component, the differences in recommendation quality observed in our experiments were relatively small. Therefore, our observations suggest that – at least for the recommendation scenarios considered in this work – all three simple inter-item proximity matrices present good candidates for high quality recommendations, with the $\mathbf{K}_{\mathbf{cos}}$ being slightly more convenient to handle computationally.

## 5 Evaluation of the computational performance in distributed memory environments

In this section we test the computational performance of our method in distributed memory environments. The experiments were performed on the `Mesabi` Linux cluster at the Minnesota Supercomputing Institute. `Mesabi` is an HP Linux cluster with a total of 741 nodes of various configurations and a total of 17,784 compute cores provided by Intel Haswell E5-2680v3 processors. Each standard node of `Mesabi` features 64 GB of RAM. In total, `Mesabi` provides 711 Tflop/s of peak performance.

We implemented EigenRec in Fortran 90. Communication among the set of available processors was performed by means of the Message Passing Interface (MPI) standard [36], and each MPI process was tied to a distinct compute core of `Mesabi`. Moreover, for each MPI process we set the number of threads equal to one. The source codes were compiled with the Intel MPI compiler (`mpiifort`) using the -O3 optimization level, and all computations were performed in 64-bit arithmetic. The linear algebra operations were performed by the appropriate routines in the Intel Math Kernel (Release 11.3) scientific library [6].

Table 2 reports the speedups (for up to 64 MPI processes) of the distributed memory implementation of Lanczos over its sequential execution, as well as the total number of iterations performed by Lanczos to compute the $f = 50, 100, 150, 200, 300$ leading eigenvectors of $\mathbf{A} = \mathbf{W}^{\intercal}\mathbf{W}$ for the complete `MovieLens20M` and `Yahoo` datasets. As we increase the number of MPI processes, the scalability of EigenRec is controlled by two key factors: a) the intrinsic sparsity of the recommender datasets, and b) the operations performed within the Lanczos algorithm. The sparsity of the datasets shifts the MV product to be more memory-bound (limited by bandwidth), in the sense that the CPU is not fully exploited each time we read data from the memory (a reality inherent to all methods based on sparse matrix computations). Moreover, the rest of the computations performed in parallel – like the inner products and the orthogonalization procedure – are generally dominated by latency and low granularity, which in turn also puts a limitation to the maximum scalability of the method[6].

[6] Note that to alleviate this, one can use sophisticated parallel schemes that try to overlap communication with computations; however, their analysis goes deep into high-performance computing and lies outside the scope of this paper.

**Table 2** Lanczos speedups over sequential execution for an increasing number of MPI processes (also shown graphically in the last row). '*Lanczos steps*': total number of iterations performed by Lanczos

| | MovieLens20M | | | | | Yahoo | | | | |
|---|---|---|---|---|---|---|---|---|---|---|
| | f=50 | 100 | 150 | 200 | 300 | f=50 | 100 | 150 | 200 | 300 |
| *1 core* | 1.00 | 1.00 | 1.00 | 1.00 | 1.00 | 1.00 | 1.00 | 1.00 | 1.00 | 1.00 |
| *2 cores* | 1.76 | 1.97 | 1.72 | 1.90 | 1.65 | 2.00 | 2.32 | 2.35 | 2.40 | 1.93 |
| *4 cores* | 3.20 | 3.44 | 2.96 | 3.20 | 2.71 | 3.45 | 3.71 | 3.77 | 3.86 | 3.15 |
| *8 cores* | 5.40 | 5.62 | 4.73 | 5.05 | 4.10 | 4.32 | 4.75 | 5.16 | 5.19 | 4.01 |
| *16 cores* | 8.38 | 8.58 | 7.02 | 7.23 | 5.80 | 6.52 | 6.97 | 7.88 | 7.79 | 5.99 |
| *32 cores* | 11.57 | 11.50 | 9.46 | 9.40 | 7.27 | 8.78 | 9.30 | 9.79 | 9.57 | 8.94 |
| *64 cores* | 15.19 | 16.32 | 13.35 | 13.44 | 10.34 | 10.50 | 11.29 | 10.42 | 11.23 | 10.66 |
| *Lanczos steps* | 160 | 290 | 430 | 570 | 790 | 150 | 280 | 410 | 540 | 820 |
| EigenRec | 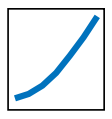 | 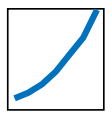 | 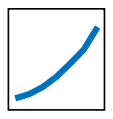 | 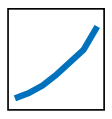 | 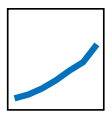 | 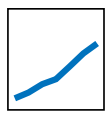 | 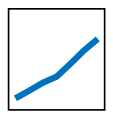 | 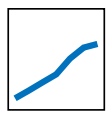 | 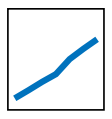 | 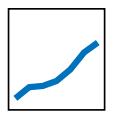 |

Taking into account these inherent restrictions of the underlying problem, we can see that the scalability of the method is satisfactory.

Figure 6 plots the percentage of the total wall-clock times of the EigenRec computation spent on MV products (denoted "MV"), and orthogonalization (denoted by "Orth"), for the experiments reported in Table 2. As we increase the number of sought latent factors $f$, the percentage of time spent on orthogonalization increases as well. Note here that the complexity of the orthogonalization procedure is quadratic to the number of iterations in Lanczos, while that of MV only linear. Thus, as the value of $f$ increases, we expect orthogonalization to account for a higher percentage of the wall-clock times, especially for datasets for which the ratio of *nnz* over the number of items is small. Indeed, for the `Yahoo` dataset, the latter ratio is more than 5250, and, for the values of $f$ tested, performing the MV products requires far more time than the orthogonalization procedure. On the other hand, for the `MovieLens20M` dataset, for which this ratio is less that 750, the percentage of the amount of time spent on orthogonalization is considerable even for $f = 50$, and it reaches to more than 50% when we use 32 MPI processes to compute $f = 300$ latent factors. In addition to the increase of orthogonalization costs for higher values of $f$, we can also notice an increase in the percentage of time spent on orthogonalization as the number of MPI processes increases. Different options to decrease the amount of time spent on orthogonalization is to combine Lanczos with polynomial filtering [2] and/or thick restarting [38]. Another alternative is to use Lanczos in combination with domain decomposition approaches [23].

Note here that in contrast with the experiments performed in §4 (where we needed to work with a small subset of `Yahoo` in order to handle the computational burden of running the competing algorithms), in this section we consider the complete `Yahoo` dataset, which contains around 717 million ratings given by over 1.8 million users to more than 136 thousand songs. Even for such large-scale problem, our parallel implementation of EigenRec exploiting no more that 64 cores, allows us to compute the recommendations very efficiently. For example if we fix $d$, $f$ to the values that give the best results in our qualitative tests, computing recommendations for all users takes around 5 seconds for the `MovieLens20M` and less than 6 minutes for the `Yahoo` dataset.

## 6 Remarks on related work

Factorization of a sparse similarity matrix was used to predict ratings of jokes in the Eigen-Taste system [18]. The authors first calculate the Pearson's correlation scores between the jokes and then form a denser latent space in which they cluster the users. The predicted rating of a user about a particular item is then calculated as the mean rating of this item, made by the rest of the users in the same cluster. The approach followed here differs significantly. The fact that we pursue ranking-based recommendations grants us the flexibility of not caring about the exact recommendation scores and allows us to introduce our novel proximity matrix, which except its pure similarity core also includes an important scaling component which was found to greatly influence the overall quality in every recommendation scenario.

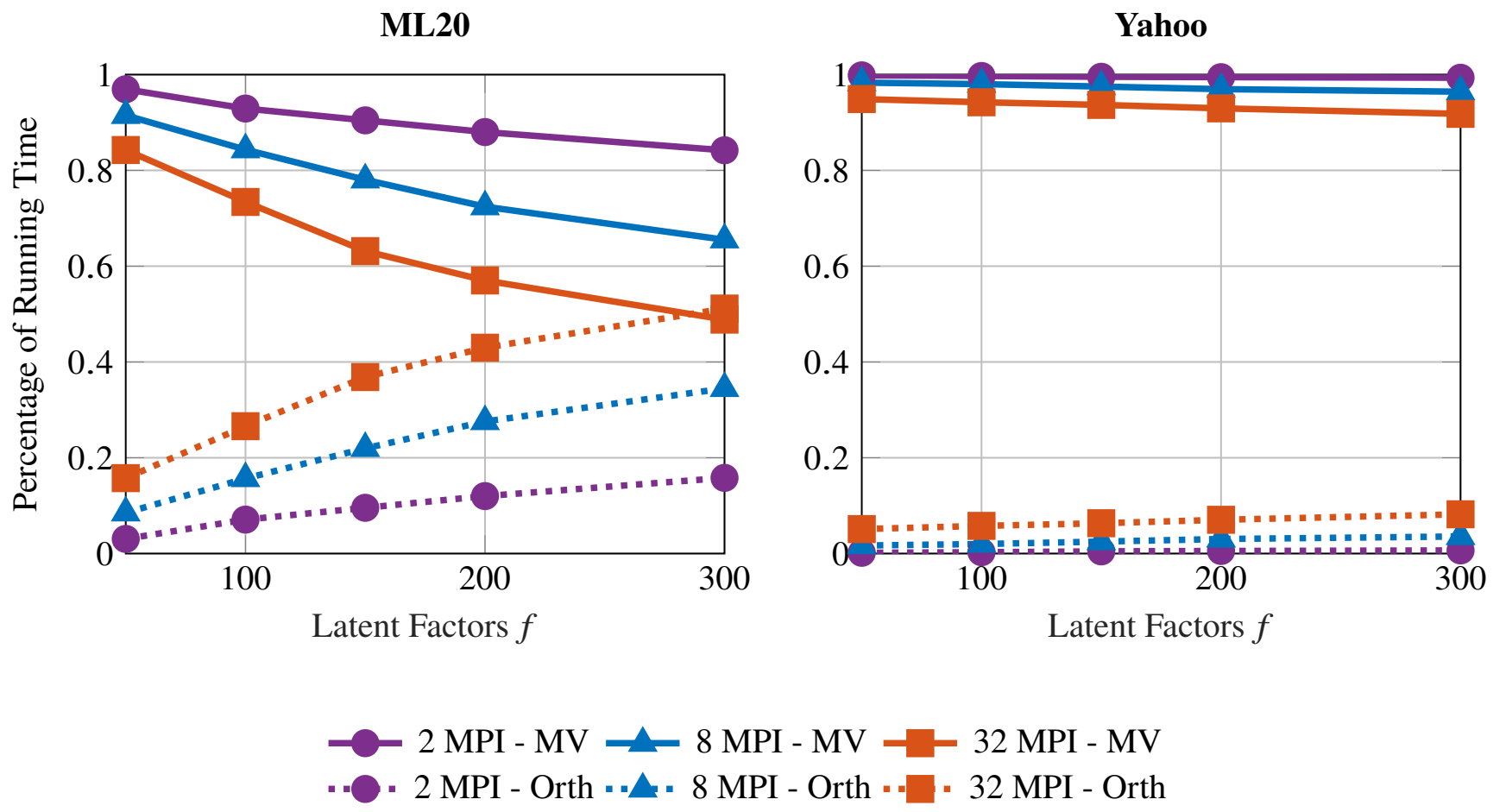

**Fig. 6** Percentage of the total wall-clock time of Lanczos spent on MV (left) and orthogonalization (right) for the experiments reported in Table 2

In the literature one can find many "content-aware" methods (both learning-based [17, 13, 35] and graph-based [27, 31]) that deal with cold-start problems exploiting meta-information about the items and/or the users outside the ratings matrix (e.g. the genre or the director of a movie, the composer of a piece of music etc.). EigenRec, on the contrary, is a pure collaborative filtering method, i.e. it neither assumes nor exploits any information about the users or the items other than the ratings matrix.

The computational core of our method is the classic Lanczos algorithm, which, together with his modifications, has been extensively used in the context of numerical linear algebra for the computation of the eigenvectors and/or singular triplets of large sparse matrices[7]. From a qualitative perspective, Blom and Ruhe [7] suggested the use of an algorithm closely related to Latent Semantic Indexing, which employs the Lanczos bidiagonalization technique to generate two sets of vectors that essentially replace the left and right singular vectors, lowering the computational cost. Chen and Saad [10] have recently examined the use of Lanczos vectors in applications where the major task can be reduced to computing a matrix-vector product in the principal singular directions of the data matrix; they demonstrated the effectiveness of this approach on two different problems originated from information retrieval and face recognition. Also, in [28], the authors examine the use of Lanczos vectors for a very fast "crude" construction of a latent space that avoids overfitting extremely sparse datasets.

[7] Different approaches to compute partial singular value decompositions of sparse matrices can be found in [39, 40].

## 7 Conclusions and future work

In this work, we introduced EigenRec; a versatile and computationally efficient latent factor framework for top-$N$ recommendations; EigenRec works by building a low-dimensional subspace of a novel inter-item proximity matrix consisting of a similarity and a scaling component. We showed that the well-known PureSVD algorithm can be seen within our framework and we demonstrated experimentally that its implicit suboptimal treatment of the prior popularity of the items inevitably limits the quality of the recommendation lists it yields; a problem that can be painlessly alleviated through our approach.

An interesting direction that we are currently pursuing involves the definition of richer inter-item proximity matrices, and the exploration of their effect in recommendation quality. The scaling component in particular, could be exploited to account for the fact that virtually all data for training recommender systems are subject to selection biases [21, 33] (e.g. one could define this component in a systematic way, incorporating information about the propensities of observing the data). In this paper, we restricted ourselves in using simple components that can be handled efficiently from a computational point of view while being able to yield good recommendations. We performed a comprehensive set of experiments on real datasets and we showed that EigenRec achieves very good results in widely used metrics against several state-of-the-art graph-based collaborative filtering techniques. Our method was also found to behave particularly well even when the sparsity of the dataset is severe – as in the New Community and the New Users versions of the Cold-Start problem – where it outperformed all other methods considered. Finally, our experiments suggest that EigenRec has a favorable computational profile and presents a viable candidate for big data scenarios.

**Acknowledgements** Vassilis Kalantzis was partially supported by a Gerondelis Foundation Fellowship. The authors acknowledge the Minnesota Supercomputing Institute (http://www.msi.umn.edu) at the University of Minnesota for providing resources that contributed to the research results reported within this paper.

## References

1. Anderson, C.: The long tail: Why the future of business is selling less of more. Hyperion (2008)
2. Aurentz, J.L., Kalantzis, V., Saad, Y.: cucheb: A gpu implementation of the filtered lanczos procedure. Computer Physics Communications (2017)
3. Baeza-Yates, R., Ribeiro-Neto, B.: Modern Information Retrieval, 2nd edn. Addison-Wesley Publishing Company, USA (2008)
4. Bai, Z., Demmel, J., Dongarra, J., Ruhe, A., van der Vorst, H.: Templates for the solution of algebraic eigenvalue problems: a practical guide, vol. 11. Siam (2000)
5. Balakrishnan, S., Chopra, S.: Collaborative ranking. In: Proceedings of the fifth ACM international conference on Web search and data mining, WSDM '12, pp. 143–152. ACM, New York, NY, USA (2012). DOI 10.1145/2124295.2124314. URL `http://doi.acm.org/10.1145/2124295.2124314`
6. Blackford, L.S., Petitet, A., Pozo, R., Remington, K., Whaley, R.C., Demmel, J., Dongarra, J., Duff, I., Hammarling, S., Henry, G., et al.: An updated set of basic linear algebra subprograms (blas). ACM Transactions on Mathematical Software **28**(2), 135–151 (2002)
7. Blom, K., Ruhe, A.: A krylov subspace method for information retrieval. SIAM Journal on Matrix Analysis and Applications **26**(2), 566–582 (2004)
8. Bobadilla, J., Ortega, F., Hernando, A., GutiéRrez, A.: Recommender systems survey. Know.-Based Syst. **46**, 109–132 (2013). DOI 10.1016/j.knosys.2013.03.012. URL `http://dx.doi.org/10.1016/j.knosys.2013.03.012`
9. Chebotarev, P., Shamis, E.: The matrix-forest theorem and measuring relations in small social groups. Automation and Remote Control **58**(9), 1505–1514 (1997)
10. Chen, J., Saad, Y.: Lanczos vectors versus singular vectors for effective dimension reduction. Knowledge and Data Engineering, IEEE Transactions on **21**(8), 1091–1103 (2009)

11. Cremonesi, P., Koren, Y., Turrin, R.: Performance of recommender algorithms on top-n recommendation tasks. In: Proceedings of the fourth ACM conference on Recommender systems, RecSys '10, pp. 39–46. ACM (2010). DOI 10.1145/1864708.1864721. URL http://doi.acm.org/10.1145/1864708.1864721
12. Desrosiers, C., Karypis, G.: A comprehensive survey of neighborhood-based recommendation methods. In: F. Ricci, L. Rokach, B. Shapira, P.B. Kantor (eds.) Recommender Systems Handbook, pp. 107–144. Springer US (2011). DOI 10.1007/978-0-387-85820-3_4. URL http://dx.doi.org/10.1007/978-0-387-85820-3_4
13. Elbadrawy, A., Karypis, G.: User-specific feature-based similarity models for top-n recommendation of new items. ACM Trans. Intell. Syst. Technol. **6**(3), 33:1–33:20 (2015). DOI 10.1145/2700495. URL http://doi.acm.org/10.1145/2700495
14. Fouss, F., Francoisse, K., Yen, L., Pirotte, A., Saerens, M.: An experimental investigation of kernels on graphs for collaborative recommendation and semisupervised classification. Neural Networks **31**, 53–72 (2012)
15. Fouss, F., Pirotte, A., Renders, J., Saerens, M.: Random-walk computation of similarities between nodes of a graph with application to collaborative recommendation. Knowledge and Data Engineering, IEEE Transactions on **19**(3), 355–369 (2007)
16. Gallopoulos, E., Philippe, B., Sameh, A.H.: Parallelism in Matrix Computations, 1st edn. Springer Publishing Company, Incorporated (2015)
17. Gantner, Z., Drumond, L., Freudenthaler, C., Rendle, S., Schmidt-Thieme, L.: Learning attribute-to-feature mappings for cold-start recommendations. In: Data Mining (ICDM), 2010 IEEE 10th International Conference on, pp. 176–185 (2010). DOI 10.1109/ICDM.2010.129
18. Goldberg, K., Roeder, T., Gupta, D., Perkins, C.: Eigentaste: A constant time collaborative filtering algorithm. Information Retrieval **4**(2), 133–151 (2001). DOI 10.1023/A:1011419012209. URL http://dx.doi.org/10.1023/A%3A1011419012209
19. Harper, F.M., Konstan, J.A.: The movielens datasets: History and context. ACM Trans. Interact. Intell. Syst. **5**(4), 19:1–19:19 (2015). DOI 10.1145/2827872. URL http://doi.acm.org/10.1145/2827872
20. Huang, Z., Chen, H., Zeng, D.: Applying associative retrieval techniques to alleviate the sparsity problem in collaborative filtering. ACM Trans. Inf. Syst. **22**(1), 116–142 (2004). DOI 10.1145/963770.963775. URL http://doi.acm.org/10.1145/963770.963775
21. Joachims, T., Swaminathan, A., Schnabel, T.: Unbiased learning-to-rank with biased feedback. In: Proceedings of the Tenth ACM International Conference on Web Search and Data Mining, WSDM '17, pp. 781–789. ACM, New York, NY, USA (2017). DOI 10.1145/3018661.3018699. URL http://doi.acm.org/10.1145/3018661.3018699
22. Kabbur, S., Ning, X., Karypis, G.: Fism: Factored item similarity models for top-n recommender systems. In: Proceedings of the 19th ACM SIGKDD International Conference on Knowledge Discovery and Data Mining, KDD '13, pp. 659–667. ACM, New York, NY, USA (2013). DOI 10.1145/2487575.2487589. URL http://doi.acm.org/10.1145/2487575.2487589
23. Kalantzis, V., Li, R., Saad, Y.: Spectral schur complement techniques for symmetric eigenvalue problems. Electronic Transactions on Numerical Analysis **45**, 305–329 (2016)
24. Koren, Y.: Factorization meets the neighborhood: A multifaceted collaborative filtering model. In: Proceedings of the 14th ACM SIGKDD International Conference on Knowledge Discovery and Data Mining, KDD '08, pp. 426–434. ACM, New York, NY, USA (2008). DOI 10.1145/1401890.1401944. URL http://doi.acm.org/10.1145/1401890.1401944
25. Koren, Y., Bell, R., Volinsky, C.: Matrix factorization techniques for recommender systems. Computer **42**(8), 30–37 (2009)
26. Lanczos, C.: An iteration method for the solution of the eigenvalue problem of linear differential and integral operators. United States Governm. Press Office (1950)
27. Nikolakopoulos, A., Garofalakis, J.: NCDREC: A decomposability inspired framework for top-n recommendation. In: Web Intelligence (WI) and Intelligent Agent Technologies (IAT), 2014 IEEE/WIC/ACM International Joint Conferences on, vol. 1, pp. 183–190 (2014). DOI 10.1109/WI-IAT.2014.32
28. Nikolakopoulos, A.N., Kalantzi, M., Garofalakis, J.D.: On the use of lanczos vectors for efficient latent factor-based top-n recommendation. In: Proceedings of the 4th International Conference on Web Intelligence, Mining and Semantics (WIMS14), WIMS '14, pp. 28:1–28:6. ACM, New York, NY, USA (2014). DOI 10.1145/2611040.2611078. URL http://doi.acm.org/10.1145/2611040.2611078
29. Nikolakopoulos, A.N., Kalantzis, V., Gallopoulos, E., Garofalakis, J.D.: Factored proximity models for top-n recommendations. In: 2017 IEEE International Conference on Big Knowledge (ICBK), pp. 80–87 (2017). DOI 10.1109/ICBK.2017.14
30. Nikolakopoulos, A.N., Korba, A., Garofalakis, J.D.: Random surfing on multipartite graphs. In: 2016 IEEE International Conference on Big Data (Big Data), pp. 736–745 (2016). DOI 10.1109/BigData.2016.7840666

31. Nikolakopoulos, A.N., Kouneli, M.A., Garofalakis, J.D.: Hierarchical itemspace rank: Exploiting hierarchy to alleviate sparsity in ranking-based recommendation. Neurocomputing **163**, 126–136 (2015). DOI 10.1016/j.neucom.2014.09.082. URL http://dx.doi.org/10.1016/j.neucom.2014.09.082
32. Sarwar, B., Karypis, G., Konstan, J., Riedl, J.: Application of dimensionality reduction in recommender system-a case study. Tech. rep., DTIC Document (2000)
33. Schnabel, T., Swaminathan, A., Singh, A., Chandak, N., Joachims, T.: Recommendations as treatments: Debiasing learning and evaluation. In: Proceedings of the 33rd International Conference on International Conference on Machine Learning - Volume 48, ICML'16, pp. 1670–1679. JMLR.org (2016). URL http://dl.acm.org/citation.cfm?id=3045390.3045567
34. Shani, G., Gunawardana, A.: Evaluating recommendation systems. In: F. Ricci, L. Rokach, B. Shapira, P.B. Kantor (eds.) Recommender Systems Handbook, pp. 257–297. Springer US (2011). DOI 10.1007/978-0-387-85820-3_8. URL http://dx.doi.org/10.1007/978-0-387-85820-3_8
35. Sharma, M., Zhou, J., Hu, J., Karypis, G.: Feature-based factorized bilinear similarity model for cold-start top-n item recommendation. In: Proceedings of the 2015 SIAM International Conference on Data Mining, SDM '15, pp. 190–198 (2015). DOI 10.1137/1.9781611974010.22. URL http://epubs.siam.org/doi/abs/10.1137/1.9781611974010.22
36. Snir, M., Otto, S., Huss-Lederman, S., Walker, D., Dongarra, J.: MPI-The Complete Reference, Volume 1: The MPI Core, 2nd. (revised) edn. MIT Press, Cambridge, MA, USA (1998)
37. Takács, G., Pilászy, I., Németh, B., Tikk, D.: Scalable collaborative filtering approaches for large recommender systems. J. Mach. Learn. Res. **10**, 623–656 (2009). URL http://dl.acm.org/citation.cfm?id=1577069.1577091
38. Wu, K., Simon, H.: Thick-restart lanczos method for large symmetric eigenvalue problems. SIAM Journal on Matrix Analysis and Applications **22**(2), 602–616 (2000)
39. Wu, L., Romero, E., Stathopoulos, A.: Primme_svds: A high-performance preconditioned svd solver for accurate large-scale computations. arXiv preprint arXiv:1607.01404 (2016)
40. Wu, L., Stathopoulos, A.: A preconditioned hybrid svd method for accurately computing singular triplets of large matrices. SIAM Journal on Scientific Computing **37**(5), S365–S388 (2015)
41. Yahoo Webscope Program: Yahoo!R2Music Dataset. https://webscope.sandbox.yahoo.com/
42. Yin, H., Cui, B., Li, J., Yao, J., Chen, C.: Challenging the long tail recommendation. Proceedings of the VLDB Endowment **5**(9), 896–907 (2012)
43. Zhou, D., Bousquet, O., Lal, T.N., Weston, J., Schölkopf, B.: Learning with local and global consistency. In: S. Thrun, L.K. Saul, B. Schölkopf (eds.) Advances in Neural Information Processing Systems 16 [Neural Information Processing Systems, NIPS 2003, December 8-13, 2003, Vancouver and Whistler, British Columbia, Canada], pp. 321–328. MIT Press (2003). URL http://papers.nips.cc/paper/2506-learning-with-local-and-global-consistency